\documentclass[12pt]{iopart}
\usepackage{epsfig,rotating,graphics}

\def\be{\begin{eqnarray} &&} 
\def\nonu{\nonumber \\ &&} 
\def\ee{\end{eqnarray}} 

\def\sumint{\int \! \!\ \! \! \! \! \!\ \! \! \!\! \!\sum}
\newcommand{\blf}[1]{\bf  {\tilde #1}}
\newcommand{\bm}[1] {\mbox{\boldmath{$#1$}}}
\newcommand{\bfi}{\begin{figure}}
\newcommand{\efi}{\end{figure}}

\begin{document}

\title[EMC effect, few-nucleon systems and Poincar\'e covariance]{EMC effect, few-nucleon systems and
 Poincar\'e covariance}

\author{Emanuele Pace$^1$,  Matteo Rinaldi$^2$, Giovanni Salm\`e$^3$ and Sergio Scopetta$^2$}

\address{$^1$Universit\`a di Roma ``Tor Vergata'' and INFN, Sezione di Roma Tor Vergata, Italy}
\address{$^2$Universit\`a di Perugia and INFN, Sezione di Perugia, Italy}
\address{$^3$INFN, Sezione di Roma, Italy}
\ead{pace@roma2.infn.it}
\vspace{10pt}
\begin{indented}
\item[]February 2020
\end{indented}

\begin{abstract}
An approach for a Poincar\'e covariant description of nuclear structure and of lepton scattering off nuclei is proposed within the relativistic Hamiltonian dynamics in the light-front form. Indeed a high level of accuracy is needed for a  comparison with the increasingly precise present and future experimental data at high momentum transfer. Therefore,  to distinguish genuine QCD effects or effects of medium modified nucleon structure functions  from conventional nuclear structure effects, the commutation rules between the Poincar\'e generators should be satisfied. For the first time in this paper
 a proper hadronic tensor for inclusive deep inelastic scattering of electrons off nuclei is  derived in  the impulse approximation in terms of the single nucleon  hadronic tensor.  Our approach is based : i) on a  light-front spectral function for nuclei, obtained taking advantage of the  successful non-relativistic knowledge of nuclear interaction, and ii) on the free current operator that, if defined in the Breit reference frame with the momentum transfer, $\bf q$, parallel to the $z$ axis, fulfills Poincar\'e covariance and current conservation.
  Our results can be generalized : i) to exclusive processes or to semi-inclusive deep inelastic scattering processes; ii) to the case where the final state interaction is considered through a Glauber approximation; iii) to finite momentum transfer kinematics. 
 As a first test, the hadronic tensor is applied to  obtain the nuclear structure function F$_2^A$ and to evaluate the EMC effect for $^3He$ in the Bjorken limit. 
 Encouraging 
 results including only the two-body part of the light-front spectral function are presented.
 
 \end{abstract}


\submitto{\PS}

\section{Introduction}
\label{intro}

The picture of nuclei as bound systems of nucleons and pions within a non-relativistic framework has reached a wide success in describing the properties of nuclei, although the fundamental theory for strong interacting systems is QCD with quarks and gluons as elementary degrees of freedom.

One of the most challenging issues is to understand  the dynamical mechanism that is able to confine quarks and gluons in the effective degrees of freedom investigated by nuclear physics, i.e. meson and nucleon.
This issue is nowadays one of the main  motivations of  experimental efforts of an increasing worldwide community, like the initiatives for the development of  the Electron Ion Collider (EIC) (see, e.g., \cite{Accardi:2012qut,Aschenauer:2014cki,Abeyratne:2012ah,Meziani}). In view of this, also theoretical efforts aiming at improving  the description of nuclei are highly desirable, since after all nuclei represent the best QCD laboratory that Nature might yield.  It is very important to  take into account as many general principles as we can,  particularly the Poincar\'e covariance, since it appears to be a prerequisite for avoiding possible ambiguities in the assessment of different effects.

A phenomenon that could indicate an  explicit  manifestation of QCD degrees of freedom in nuclei is the EMC effect  \cite{EMC}, which points to modifications of the nucleons bound in the nuclear medium \cite{Cloet}.
However, to clearly isolate genuine QCD effects in nuclei the general relativistic constraints, as the proper commutation rules between the Poincar\'e generators, should be satisfied 
 when processes involving nucleons with 
high 3-momentum are considered and a high
precision is needed. For instance this is the case for the 
  JLab program at 12 GeV, both for the inclusive and the semi-inclusive deep inelastic processes (see, e.g., \cite{JLAB} and \cite{SIDIS1,SIDIS2,SIDIS3}, respectively) or for the future experiments to be performed at the EIC \cite{Accardi:2012qut,Aschenauer:2014cki,Abeyratne:2012ah}.

{Traditional nuclear physics has achieved a deep knowledge of the nuclear interaction and has developed sophisticated methods for evaluating the bound state wave functions of nuclei in a non-relativistic framework.
Our work points to merge this knowledge }
in a Poincar\'e covariant framework through the light-front (LF) form of the relativistic Hamiltonian dynamics of an interacting system \cite{Dirac,KP}, where the LF components $[v^-,\tilde{\bm v}]=[v^-, v^+,{\bm v}_\perp]$ of a four vector $v$ are used, with $v^{\pm}= v_0 \pm v_z$. Indeed, within the relativistic Hamiltonian dynamics plus the Bakamjan-Thomas construction of the Poincar\'e generators \cite{BT}, covariance of the Poincar\'e generators imposes to the interaction the same commutation rules as  occurs for the non-relativistic interaction and then the phenomenological interactions obtained from the analysis of the phase shifts can be used in a  Poincar\'e covariant environment (see, e.g., \cite{noi}). Furthermore LF dynamics allows for a meaningful Fock state expansion of wavefunctions \cite{Brodsky}.

The tool we use  is the LF spectral function \cite{noi}, which allows one to take advantage of the whole successful non-relativistic phenomenology for the nuclear interaction  and to take care of the macroscopic locality 
(for a definition of macroscopic locality see, e.g., \cite{KP}). As shown in \cite{noi}, from the LF spectral function one obtains a momentum distribution which fulfills at the same time the baryon number sum rule and the momentum sum rule. 

The process we intend to study is the {{EMC}} effect, which gives evidence that the structure of nucleons may be different  when bound together in nuclei and has been and still is the subject of extensive experimental programs in many laboratories (see, e.g.,  \cite{Cloet}, \cite{JLAB} and references quoted therein).
To this end we  first  obtain the proper LF expressions for the hadronic tensor and for the $F_2$ nuclear structure functions in the impulse approximation. After that, the obtained expression for $F_2$ is applied to the EMC effect for {$^3He$} (for some preliminary results see \cite{prel}).

In this paper, after a short review of the {{EMC}} effect in Section 2, the definition of the LF spectral function 
is recalled in Section 3. Then in Sections 4 and 5 it is shown for the first time for a generic $A$-nucleon nucleus how the hadronic tensor and  the nuclear structure function $F_2$ can be obtained  within 
 LF Hamiltonian dynamics  through the LF spectral function in the Bjorken limit. In Section 6  {{preliminary results}} for the {EMC effect in {$^3He$} with the LF spectral function} are presented and {our
conclusions} are drawn in Section 7.

\section{A short review of the {{EMC}} effect}
\label{EMC}

The EMC effect \cite{EMC} was observed in deep inelastic lepton scattering (DIS) experiments  
 and is described by the ratio of the DIS cross section on an $A$-nucleus and on the deuteron
\begin{eqnarray}
R(x) = {2 \sigma_A \over A \sigma_D}\ne 1 \hspace{2cm}
{\rm{with}} 
\hspace{2cm}
 x = {Q^2 \over 2m\nu}  
 \end{eqnarray}
in the range 0 $\le  x \le$ 1, where 
$m$ is the nucleon mass, $Q^2 = - q^2$, $q$ the four-momentum transfer and $\nu$  the  energy transfer from the lepton to the nucleus in the laboratory frame. It suggests that a picture of lepton scattering off nuclei as 
an incoherent sum of lepton scattering off the constituent protons 
and  neutrons is incomplete and
 provides 
 an indication for {{explicit QCD effects
 in nuclei}}.
Since the original discovery, a large program of measurements at CERN, Fermilab, SLAC, DESY, and JLab aimed at understanding the EMC effect.

Following \cite{Cloet}, we report here the main {{{experimental conclusions : i) the shape of the effect is universal and observed in all nuclei; ii) it is relatively $Q^2$ independent and slowly increases with $A$; iii) the $A$ dependence is consistent with a dependence upon the local nuclear density.}} 

A linear relation was found  \cite{Arr} between the size of the EMC effect measured as the slope of $R(x)$ for 0.35 $\le  x \le$ 0.7
 and the short range correlations (SRCs) taken as the ratio 
\begin{eqnarray}
\hspace{0.5cm} a_2 = {\sigma_A \over  \sigma_D} \hspace{2mm} 
\hspace{2.5cm}
{\rm{at }} 
\hspace{2.5cm}
 x > 1   \quad .
  \end{eqnarray}
However, it is unclear whether the SRCs cause the 
EMC effect or if there is some common  underlying source of the two phenomena.

{ {
There are two categories for the {{theoretical models}} that have been proposed : i) the first one takes care only of  {{traditional nuclear physics}}, including binding and nucleon momentum distributions effects through spectral functions corresponding to realistic nuclear interactions; ii) 
the second one invokes {{more exotic explanations}}, such as contributions of six or nine quark bags, or medium modification of the internal structure of nucleons  bounded in a nucleus. }}

 {Our final aim is  to provide 
 {{{a LF Poincar\'e covariant calculation of the nuclear part of the EMC effect for  $^3He$ and $^3H$}}}}
as a solid starting point which allows one to safely investigate genuinely QCD-based effects.

\section{The LF spectral function
 }
\label{SF}

{{The spectral function}} is the {{probability distribution to find inside a bound system
 a particle with a given
3-momentum when the rest of the system has a given energy $\epsilon$.} 
}

The definition of the  spin dependent LF spectral function of a nucleon in a nucleus 
   of total 4-momentum $P_A$ in the laboratory frame, with LF momentum components 
  $[P_A^-, {\blf P}_A] \equiv [P_A^-, P_A^+,{\bf P}_{A\perp}]$, is based on the overlaps
\be
_{LF}\langle  t T ; 
\alpha,\epsilon ;J J_{z}; \tau\sigma,\tilde{\bm \kappa}|
j,j_z=m;
\epsilon^A,\Pi; T_A T_{Az}\rangle  \quad .
\label{overlap}
\ee
In (\ref{overlap})
the state
$|j,j_z=m; \epsilon^A, \Pi; T_A T_{Az}\rangle$ is the intrinsic ground eigenstate of the $A$-nucleon system, with angular momentum $(j,j_z)$, energy $\epsilon^A$, parity $\Pi$, isospin  $(T_A T_{Az})$, and
polarized along $\hat z$,  with LF total energy $P_A^-$.
The quantum numbers ($J, J_{z}; 
\epsilon;  T t$) describe the angular momentum, intrinsic energy and isospin, respectively, of the $(A-1)$-nucleon system, while $\alpha$ indicates the other quantum numbers needed to completely identify this system.
 The intrinsic state $\left |\tilde{\bm \kappa}, \sigma \tau; J J_{z} 
\epsilon, \alpha, T t\right\rangle_{LF}$ is a cluster composed by 
a fully-interacting intrinsic state of ~$(A-1)$ nucleons and
 a plane wave, describing a nucleon that freely
moves in the intrinsic reference frame of the whole cluster $[1, (A-1)]$ 
with intrinsic momentum
\be
\kappa^+={\cal{M}}_0[1,(A-1)] ~ \xi   
\nonu
{\bm \kappa}_{\perp}= {\bf p_{\perp}}-\xi ~{\bf P}_{A\perp} \quad .
\label{intr}
\ee
Let us stress that the intrinsic momentum $\tilde{\bm \kappa}\equiv (\kappa^+,{\bm \kappa}_{\perp})$ of a nucleon in the reference frame of the  cluster $[1, (A-1)]$ is different from the momentum $\tilde{\bm p} \equiv ( p^+,{\bf p}_{\perp})$ in the laboratory frame.
 As noticed in \cite{KP,noi}, the cluster $[1, (A-1)]$ fulfills the macrocausality \cite{KP,noi}. 
In  (\ref{intr}) one has $\xi = p^+/P^+_A$ and ${\cal{M}}_0[1,(A-1)]$ is the intrinsic energy of the cluster
\be
\fl  {\cal{M}}_0[1,(A-1)]= {(\kappa^+)^2 + (m^2 + k^2_{\perp}) \over 2 ~ \kappa^+ }
 ~+\left [ \left [(\kappa^+)^2 + (m^2 + k^2_{\perp}) \over 2 ~ \kappa^+ \right ]^2 ~ +
 ~ M_S^2 ~ - ~ m^2 \right ]^{1/2} \quad .
\label{M0}
\ee
where $M_S$ is the mass of the interacting spectator $(A-1)$-nucleon system.

  For an $A$-nucleon 
  system of total 4-momentum $P_A$ 
 and
   polarized  along the {{polarization vector {\bm S}}}, the spin dependent LF spectral function of a nucleon with 4-momentum $p \equiv [p^-, p^+,{\bf p}_{\perp}]$ in the laboratory frame
 can be defined as follows (see (69) of  \cite{noi})

\be
\fl 
{\cal P}^{\tau}_{\sigma'\sigma}(\tilde{\bm \kappa},\epsilon,{\bf S},{\cal M}) = 
 \sum_{m,m'}  ~ D^{{j}*}_{m, \cal M}(\alpha, \beta, \gamma) ~
D^{j}_{m', \cal M}(\alpha, \beta, \gamma) ~
   {\cal P}^{\tau, m',m}_{\sigma'\sigma} (\tilde{\bm \kappa},\epsilon,\hat z)   \quad ,
\ee
where $D^{j}_{m, \cal M}(\alpha, \beta, \gamma) $ is the rotation matrix, with $\alpha, \beta, \gamma$ the Euler angles describing the proper rotation from the $z$ axis to the polarization vector {\bm S}, and 
\be
\fl  {\cal P}^{\tau, m',m}_{\sigma'\sigma}(\tilde{\bm \kappa},\epsilon,\hat z) =
\rho(\epsilon) ~\sum_{J J_{z}\alpha}\sum_{T t } ~
_{LF}\langle  t T ; 
\alpha,\epsilon ;J J_{z}; \tau\sigma',\tilde{\bm \kappa}|
j,j_z=m';
\epsilon^A,\Pi; T_A T_{Az}\rangle
\nonu
\fl \times ~
\langle T_A T_{Az}; \Pi,\epsilon^A; j,j_z=m; 
|\tilde{\bm \kappa},\sigma\tau; J J_{z}; 
\epsilon, \alpha; T t\rangle_{LF} 
\label{SFex}
\ee
with  $\rho (\epsilon)$ the energy density of the $(A-1)$-nucleon states.

In conclusion, the LF spectral function is obtained through the overlaps between the cluster $[1, (A-1)]$ and the bound state of the $A$-particle system. In \cite{noi} it is shown for a three-particle system how these overlaps can be evaluated from the  system ground state wave function and the two-particle system wave functions for the bound and the scattering states in momentum space.

Let us consider the unpolarized spectral function, independent of ${\bf S}$,
\be
\hspace{-7mm} {\cal P}^{\tau}_{\sigma'\sigma}(\tilde{\bm \kappa},\epsilon
)  
=  {1 \over 2~ j + 1} ~ \sum_{\cal M} 
{\cal P}^{\tau}_{\sigma'\sigma}(\tilde{\bm \kappa},\epsilon,{\bf S},{\cal M}) =
 {1 \over 2~ j + 1} ~ \sum_{m} {\cal P}^{\tau,m,m}_{\sigma'\sigma} (\tilde{\bm \kappa},\epsilon,\hat z)
  \quad .
\label{unpSF} 
\ee
Because of time-reversal and parity symmetries one has
\be
\sum_{m} {\cal P}^{\tau,m,m}_{\sigma'\sigma} (\tilde{\bm \kappa},\epsilon,\hat z)  
 =   \sum_{m} (-1)^{2j+1}(-1)^{m+m+\sigma'+\sigma}
\left  [ {\cal P}^{\tau, -m,-m}_{-\sigma'-\sigma} (\tilde{\bm \kappa},\epsilon,\hat z) \right ]^* 
 \nonu 
 = - \sum_{m}  (-1)^{\sigma'+\sigma}
~ {\cal P}^{\tau,m,m}_{-\sigma-\sigma'} (\tilde{\bm \kappa},\epsilon,\hat z)   \quad .
 \label{A1V}
   \ee
It is clear that  the average on the spin directions gives a vanishing result for the non-diagonal terms of the spin-dependent spectral function. 
If $~\sigma  = \sigma'~$, then
$ \sum_{m} {\cal P}^{\tau,m,m}_{\sigma\sigma} (\tilde{\bm \kappa},\epsilon,\hat z) = \sum_{m} 
~ {\cal P}^{\tau,m,m}_{-\sigma-\sigma} (\tilde{\bm \kappa},\epsilon,\hat z) $. This means that the diagonal, unpolarized spectral function is independent of 
$\sigma$. 
In conclusion the properly normalized, unpolarized  spectral function is \cite{noi}
\be
\fl  {\cal P}^{\tau}(\tilde{\bm \kappa},\epsilon)  
= ~  {1 \over 2~ j + 1} ~ \sum_{\cal M} \sum_\sigma 
{\cal P}^{\tau}_{\sigma\sigma}(\tilde{\bm \kappa},\epsilon,{\bf S},{\cal M}) ~ = ~
 {1 \over 2~ j + 1} ~
 \sum_{m}  \sum_\sigma ~
{\cal P}^{\tau,m,m}_{\sigma\sigma} (\tilde{\bm \kappa},\epsilon,\hat z)   \quad .
\label{unpol}
\ee
   
\section{Hadronic tensor in light-front Hamiltonian dynamics}
\label{HT}

Within light-front Hamiltonian dynamics (LFHD), the hadronic tensor for  inclusive lepton-nucleus scattering off an $A$-nucleon nucleus of LF state 
$ |\Psi_{0}; {\bf S}, {\cal M},T_{z};P_A\rangle_{LF}$, which is polarized along ${\bf S}$ with spin projection $\cal M$, has isospin third component $T_{Az}$ and total 4-momentum $P_A$ in the laboratory frame, is
\be
\hspace{-0.8cm} W^{\mu \nu}_A(P_A,{\bf S},{\cal M},T_{Az}, q) = {1 \over 4 \pi} \sum_X ~_{LF}\langle\Psi_{0}; {\bf S}, {\cal M},T_{Az};P_A |J^{\mu}_A(0) | X, P_X\rangle_{LF} ~ 
 \nonu
 \times ~
_{LF}\langle X, P_X |J^{\nu}_A(0) |\Psi_{0}; {\bf S}, {\cal M},T_{Az};P_A\rangle_{LF}~
(2 \pi)^4~ 
\delta^4(q + P_A -P_X)  \quad ,
\label{HT}
\ee
where 
$ | X, P_X\rangle_{LF}$ is any LF final state.

Let us assume that the virtual photon scatters incoherently from each one of the nucleons in the nucleus, so that the final state is composed of an $(A-1)$-nucleon spectator system and
the debris  produced by the virtual photon impinging on a single nucleon. 
Then, instead of 
\be 
\sum_X  ~ | X,  P_X \rangle_{LF} ~ _{LF}\langle X,  P_X  | ~ = ~ I  \quad ,
\ee
let us introduce in (\ref{HT}) the quantity 
\be
\hspace{-1.2cm}\sum_{J J_{z} \alpha} 
  \sum_{T t } ~ \sumint { \rho (\epsilon) ~ d\epsilon 
   }
  \int {d{\blf P}_{S} \over 2 (2 \pi)^3  P^+_{S}}~\sum_f  ~ \sum_{\sigma_f \tau_f} 
~ \int { d {\blf p}_f \over 2 (2 \pi)^3 p^+_f} ~
  \times \nonu 
\hspace{-.4cm} |{\blf{p}}_f,\sigma_f \tau_f \rangle_{LF}   |{\blf P}_{S};J J_{z}, \epsilon, \alpha, T t
  \rangle_{LF} 
  ~_{LF}
  \langle 
 tT , 
\alpha,\epsilon, J_{z}J;{\blf P}_S 
| _{LF} \langle {\blf{p}}_f,\sigma_f \tau_f  | ~  \quad ,
\label{compl}
\ee
where  ${\blf P}_S$ is the total LF momentum  of a fully interacting $(A-1)$-nucleon spectator system in the laboratory.
 
  The state  $|{\blf P}_{S};J J_{z}, \epsilon, \alpha, T t\rangle_{LF}$
is an eigenstate of the operator  $\hat P^-$ with eigenvalue $P_S^-$
\be
\hspace{-.5cm} \hat P^-\left|{\blf P}_{S};J J_{z}, \epsilon, \alpha, T t\right\rangle_{LF}=
~{\left[M^2_S+|{\bf P}_{S\perp}|^2\right] \over  P^+_S}~
\left|{\blf P}_{S};J J_{z}, \epsilon, \alpha, T 
t\right\rangle_{LF}~.~~~~
\label{Pmeno}\ee
 For  $(A-1)=2$, the mass of the two-body interacting spectator system is $M_S =   
[4(m^2 + m \epsilon)]^{1/2}$ \cite{noi}.

In (\ref{compl})
  $|{\blf{p}}_f,\sigma_f \tau_f \rangle_{LF}$  is the LF
   final state of the debris  produced by the scattering of the virtual photon off a nucleon, and has
  isospin third component
   $\tau_f $,  spin along the z-axis equal to $\sigma_f $, LF momentum 
${\blf{p}}_f$ in the laboratory  frame and LF energy $p_f^- = (m_f^2 + {\bm p}^2_{f \perp})/p_f^+$. The   quantity $p_f^-$ is an independent variable with respect to the components of the LF momentum ${\blf p}_f$, since the unknown mass $m_f$ of the debris can assume any positive value compatible with   conservation laws.

Then, assuming no interaction in the final state between the  $(A-1)$-nucleon system and the state 
$|{\blf{p}}_f,\sigma_f \tau_f \rangle$, one has $P_X = P_S + p_f$ and the hadronic tensor becomes

\be
\fl  W^{\mu \nu}_A = {1 \over 4 \pi}  ~\sum_{J J_{z} \alpha} 
  \sum_{T t } ~ \sumint {  \rho (\epsilon) ~ d\epsilon
   }
  \int {d{\blf P}_{S} \over 2  (2 \pi)^3  P^+_{S}}~~\sum_f  ~ \sum_{\sigma_f \tau_f} 
  \int { d {\blf p}_f \over 2 (2 \pi)^3 p^+_f} ~(2 \pi)^4
  \nonu 
\fl  
  \times ~ \delta^4(q + P_A -p_f - P_S) ~
_{LF}\langle\Psi_{0}; {\bf S},{\cal M}, T_{Az};P_A |J^{\mu}_A(0)|{\blf{p}}_f,\sigma_f \tau_f \rangle_{LF}   |{\blf P}_{S};J J_{z}, \epsilon, \alpha, T t
  \rangle_{LF} 
   \nonu
\fl  \times ~ _{LF}
  \langle 
 t T, 
\alpha,\epsilon, J_{z}J;{\blf P}_S 
| _{LF} \langle {\blf{p}}_f,\sigma_f \tau_f  | 
  J^{\nu}_A(0) |\Psi_{0}; {\bf S}, {\cal M},T_{Az};P_A\rangle_{LF} \quad .
  \label{HT1}
\ee

In impulse approximation the current of the $A$-nucleon system is
\be
J^{\mu}_A (0) = ~ \sum_N J^{\mu}_N (0) \quad .
\label{IA}
\ee
As shown in \cite{Lev}, this current satisfies Poincar\'e covariance and current conservation in the Breit frame with the momentum transfer, $\bf q$, along the $z$ axis and in any frame that can be obtained through a LF boost parallel to the $z$ axis. To be more definite let us take $\bf q$ opposite to the $z$ axis : ${\bf q} \equiv (0,0, q_z=-|{\bf q}|)$.

Let us assume that the interference terms between the contributions of different constituent nucleons yield a negligible contribution to the hadronic tensor. Then the hadronic tensor can be approximated as follows

\be
\fl W^{\mu \nu}_A 
= ~{1 \over 4 \pi}   \sum_N ~\sum_{J J_{z} \alpha} 
  \sum_{T t } ~ \sumint { \rho (\epsilon) ~ d\epsilon
   }
  \int {d{\blf P}_{S} \over 2 (2 \pi)^3  P^+_{S}}~ ~\sum_f  ~ \sum_{\sigma_f \tau_f} 
  \int { d {\blf p}_f \over 2 (2 \pi)^3 p^+_f}~~(2 \pi)^4~ 
  \nonu 
\hspace{-1.9cm} \times ~  \delta^4(q + P_A -p_f - P_S) ~ _{LF}\langle\Psi_{0}; {\bf S}, {\cal M}, T_{Az};P_A | {\blf P}_{S};J J_{z}, \epsilon, \alpha, T t
  \rangle_{LF} 
     \nonu
\hspace{-1.9cm}     \times  \sum_{\sigma' \tau'} \int { d {\blf p}' \over 2 (2 \pi)^3 p'^+} ~
   |{\blf{p}}',\sigma' \tau' \rangle_{LF} ~ _{LF} \langle {\blf{p}}',\sigma' \tau'  |
J^{\mu}_N(0)|{\blf{p}}_f,\sigma_f \tau_f \rangle_{LF}    ~
 _{LF} \langle {\blf{p}}_f,\sigma_f \tau_f  | 
  J^{\nu} _N(0)~ 
  \nonu
\hspace{-1.9cm}\times  \sum_{\sigma \tau}  \int { d {\blf p} \over 2 (2 \pi)^3 p^+} |{\blf{p}},\sigma \tau\rangle_{LF} ~ 
  _{LF} \langle {\blf{p}},\sigma \tau|
  \langle 
 t T, 
\alpha,\epsilon, J_{z}J;{\blf P}_S |\Psi_{0}; {\bf S}, {\cal M}, T_{Az};P_A\rangle_{LF} ~  \quad ,
  \label{HT2}
\ee
where the completeness for the one-nucleon momentum eigenstates $ |{\blf{p}},\sigma \tau\rangle_{LF} $ has been inserted, with $\blf p$ the nucleon LF momentum in the laboratory frame.

By considering that the LF-momentum is conserved (the interaction is contained 
only in the minus component of the momenta) and  the kinematical nature of the LF-boosts, one has (see  \ref{orth})
\be
|{\blf p},\sigma \tau \rangle_{LF}  |{\blf P} _{S}; J J_{z}, 
\epsilon, \alpha, T t\rangle_{LF}
 = 
  \nonu
  |{\blf P} = {\blf p} + {\blf P} _{S}\rangle_{LF} ~ |\tilde{\bm \kappa},\sigma \tau; J J_{z}, 
\epsilon, \alpha, T t\rangle_{LF}~\sqrt{{ E_S \over {\cal{M}}_0[1,(A-1)]}}  \quad .
\label{ortimpT} \ee
 where $E_S = \sqrt{M_S^2 + |{\bm \kappa}|^2}$,
 and
  $|{\blf P} = {\blf p} + {\blf P} _{S}\rangle_{LF}$ is the total LF momentum eigenstate of the cluster 
  $[1, (A-1)]$.

The factor $\sqrt{{ E_S / {\cal{M}}_0[1,(A-1)]}}$ in (\ref{ortimpT})
 takes care of the proper normalization
of the momentum eigenstates $|{\blf p} \rangle_{LF}$, $| {\blf P}_{S}\rangle_{LF}$ 
and $|{\blf p} + {\blf P} _{S}\rangle_{LF}$ (see \ref{orth}). 

Then one obtains
 \be
\fl _{LF}\langle \Psi_{0}; {\bf S}, {\cal M}, T_{Az};P_A|{\blf p},\sigma \tau;{\blf P} _{S}; J J_{z} ,
\epsilon, \alpha, T t\rangle_{LF}
 = 2P^+_A (2\pi)^3 ~\delta^3({\blf P}_A -{\blf P}_{S}-{\blf p})
 \nonu
\fl  \times ~ 
 _{LF}\langle \Psi_{0}; {\bf S}, {\cal M}, T_{Az}|\tilde{\bm \kappa},\sigma \tau; J J_{z}, 
\epsilon, \alpha, T t\rangle_{LF}~\sqrt{{ E_S \over {\cal{M}}_0[1,(A-1)]}}  \quad ,
\label{ort}
\ee
where 
the orthogonality of the plane waves is given by  
 $\langle{\bf \tilde P}|{\bf \tilde P'}\rangle=2P^+(2\pi)^3
\delta({\bf \tilde P}-{\bf \tilde P'})$.
Using (\ref{ort}), the hadronic tensor (\ref{HT2}) becomes

\be
\fl W^{\mu \nu}_A =
 ~{1 \over 4 \pi}  ~ \sum_N ~\sum_{\sigma \sigma'} ~\sumint {d\epsilon 
}
  \int {d{\blf P}_{S} \over (2 \pi)^3 2 P^+_{S}}~  \left [{ P_A^+  \over  p^+} \right ]^2 ~
   { E_S \over {\cal{M}}_0[1,(A-1)]}  ~\sum_f  ~ ~2 \pi ~  \rho (\epsilon)
  \nonu
\fl \times  \delta(q^- + P_A^- -p_f^- - P_S^-) ~
   \sum_{J J_{z} \alpha} 
  \sum_{T t } ~ _{LF}\langle\Psi_{0}; {\bf S},{\cal M},  T_{Az} | \tilde{\bm \kappa},\sigma' \tau; J J_{z}, 
\epsilon, \alpha, T t\rangle_{LF} ~  
\nonu
\fl \times ~_{LF} \langle \tilde{\bm \kappa},\sigma \tau; J J_{z}, 
\epsilon, \alpha, T t |\Psi_{0}; {\bf S}, {\cal M}, T_{Az}\rangle_{LF} ~
 \sum_{\sigma_f \tau_f} 
  \int { d {\blf p}_f ~ (2 \pi)^3 \over 2 (2 \pi)^3 ~ p^+_f} 
\nonu 
\fl \times ~\delta^3({\blf q} +{\blf P}_A - {\blf p}_f - {\blf P}_S) ~
  _{LF} \langle {\blf{p}},\sigma' \tau  |
J^{\mu}_N(0)|{\blf{p}}_f,\sigma_f \tau_f \rangle _{LF}   ~
    _{LF} \langle {\blf{p}}_f,\sigma_f \tau_f  | 
  J^{\nu} _N(0) |{\blf{p}},\sigma \tau\rangle_{LF} ~  \quad .
  \label{HT3}
\ee
where $|\Psi_{0}; {\bf S}, {\cal M}, T_{Az}\rangle_{LF}$ is the intrinsic state of the nucleus with LF energy $P_A^-$. In (\ref{HT3}) the equalities ${\blf p} = {\blf p}' = {\blf P}_A - {\blf P}_S$ and $\tau = \tau ' = T_{Az} - t$ have been used and the sum over $\tau$ is implicit in the sum over $N$.

Let us insert in (\ref{HT3}) the definition of the LF spin-dependent spectral function (see (69) and (66) of \cite{noi} and Section \ref{SF})
and the definition of the hadronic tensor for a single constituent
\be
w^{\mu \nu}_{N,\sigma',\sigma}({ p},{ q})=
~{1 \over 4 \pi} ~\sum_f ~ \sum_{\sigma_f \tau_f} 
  \int { d {\blf p}_f \over 2 ~(2 \pi )^3 ~ p^+_f} ~(2 \pi )^4 ~
\delta^4({q}  - {p}_f + {p}) ~ \times \nonu
  _{LF} \langle {\blf{p}},\sigma' \tau  |
J^{\mu}_N(0)|{\blf{p}}_f,\sigma_f \tau_f \rangle _{LF}   ~
    _{LF} \langle {\blf{p}}_f,\sigma_f \tau_f  | 
  J^{\nu} _N(0) |{\blf{p}},\sigma \tau\rangle_{LF} \quad  ,
\label{sHT}
\ee
where $p^-= P_A^- - P_S^-$ (i.e., the nucleon with four momentum $p$ is an off-shell nucleon) and 
$ {\blf p} =  {\blf P}_A - {\blf P}_S$.
Then the hadronic tensor can be written as follows
\be
\fl W^{\mu \nu}_A =  
\sum_{N\sigma \sigma'} ~\sumint {d\epsilon 
 }
  \int {d{\blf P}_{S} \over 2 (2 \pi)^3  P^+_{S}}~ {1 \over  \xi^2} ~
   { E_S \over {\cal{M}}_0[1,(A-1)]}  ~
 {\cal P}^{\tau}_{\sigma\sigma'}(\tilde{\bm \kappa},\epsilon,{\bf S},{\cal M})  ~ w^{\mu \nu}_{N,\sigma',\sigma}({p},{q})  \quad ,
\label{HT4}
\ee
where the isospin $\tau$  is defined by $N$.

The hadronic tensor can be expressed as an integral on the intrinsic momentum $\blf \kappa$. Since \cite{noi}
\be
{\bm P}_{S \perp} = {\bm P}_{A \perp} - {\bm p}_{\perp} =  {\bm P}_{A \perp} -
{\bm \kappa}_{\perp} - \xi ~{\bf P}_{A\perp} \\&&
P_S^+ = P_A^+ - p^+ =  P_A^+ - \xi  P_A^+
\label{PS}
\ee
one has
\be
\left | {\partial ({\bf P}_{S \perp}, P_S^+)  \over \partial ({\bm \kappa}_\perp,\xi)} \right | = P_A^+
\quad \quad .
\label{Jac}
\ee
Then using the equality \cite{noi}
\be
{\partial  \kappa ^+ \over \partial \xi}={E_S\over(1-\xi)}  \quad ,
\label{derparz}
\ee
 the hadronic tensor becomes
\be
\fl W^{\mu \nu}_A = 
~  \sum_N ~\sum_{\sigma \sigma'} ~\sumint {d\epsilon 
}
  \int {d{\bm \kappa}_{\perp} ~d\kappa ^+ ~ \over 2~(2 \pi)^3 ~  \kappa ^+}~ {1 \over  \xi} ~
    ~ 
 {\cal P}^{N}_{\sigma\sigma'}(\tilde{\bm \kappa},\epsilon,{\bf S},{\cal M})  ~ w^{\mu \nu}_{N,\sigma',\sigma}({p},{q})  \quad .
\label{HT5}
\ee

In the rest reference frame for the nucleus,  one has $~{\bm P}_{A \perp} = 0~$ and $~P_A^+ = 
P_A^- = M_A~$. 
Therefore  one obtains
\be
{\bf p}_\perp = {\bm \kappa}_{\perp}
\nonu
p^+ = \xi ~ M_A
\nonu
p^- = M_A - {\left[ M_S^2 + |{\bm \kappa}_{\perp}|^2\right] \over  M_A ~ (1 - \xi)}  \quad .
\label{pp}
\ee

Let us notice that the hadronic tensor for a single constituent depends on the energy $\epsilon$ both through the $plus$ component of the momentum $p$ and through the $minus$ component. Indeed, since $M_S$ depends on $\epsilon$, the same occurs for  ${\cal{M}}_0[1,(A-1)]$ (see  (\ref{M0})) and in turn for $\xi$
 (see  (\ref{intr})). 

If the hadronic tensor for the nucleus is averaged on the spin projections of the nucleus along ${\bf S}$,

\be
\hspace{-0.8cm}  W^{\mu \nu}_A(P_A,T_{Az},q) 
={1 \over 2~ j + 1} \sum_{\cal M} ~ W^{\mu \nu}_A(P_A,{\bf S},{\cal M},T_{Az},q) \quad,
\ee
then one obtains the hadronic tensor for an unpolarized nucleus, which is expressed through the unpolarized spectral function, independent of ${\bf S}$.

 Eventually the  hadronic tensor for an unpolarized nucleus reads
 
\be
\hspace{-.7cm} W^{\mu \nu}_A(P_A,T_{Az},q) 
= {1 \over  2} \sum_N \sum_{\sigma} ~  
\sumint {d\epsilon 
}
  \int {d{\bm \kappa}_{\perp} ~d\kappa ^+ ~ \over 2~ (2 \pi)^3 ~  \kappa ^+}~ {1 \over  \xi} 
  ~
 {\cal P}^{N}(\tilde{\bm \kappa},\epsilon)  ~ w^{\mu \nu}_{N,\sigma,\sigma}({p},{q}) \quad .
\label{HT7}
\ee 

\section{Nuclear structure function F$_2$}
\label{FF2}
In the Bjorken limit the nuclear structure function $F^A_2$ can be obtained from the  hadronic tensor for an unpolarized nucleus as follows
\be
\fl  F^A_2(x) = ~ - {1 \over 2} ~ x_A ~ g_{\mu\nu} ~ W^{\mu \nu}_A(P_A,T_{Az}) ~ =
\nonu
\nonu
\fl  = {1 \over 2} ~ \sum_N \sum_{\sigma} ~\sumint {d\epsilon 
}
  \int {d{\bm \kappa}_{\perp} ~ \over (2 \pi)^3} ~\int_{\kappa ^+_m}^{\infty} 
  {d\kappa ^+  \over 2~ \kappa ^+}~ {1 \over  \xi} 
    ~ 
 {\cal P}^{N}(\tilde{\bm \kappa},\epsilon) ~ (  - x_A )~{1 \over 2} ~ g_{\mu\nu} ~ w^{\mu \nu}_{N,\sigma,\sigma}({p},{q})
 ~ ,
\label{F2}
\ee
where 
\be
x_A ~ = ~ {Q^2 \over 2 P_A \cdot  q} \quad .
\label{Bjorken}
\ee
In (\ref{F2}) the integration limits on $\kappa^+$ are made explicit.  The minimum value for 
$ \kappa^+$ (see (\ref{intr}) and  (\ref{M0x})) is the value corresponding to $\xi_{min}$, i.e.,
\be
\kappa ^+_m ~ = ~ \xi_{min} ~ 
\left [ {m^2+|{\bm \kappa}_{\perp}|^2 \over \xi_{min}} + 
{M_S^2+|{\bm \kappa}_{\perp}|^2 \over (1-\xi_{min})} \right ]^{1/2}
\quad 
\label{kmin}
\ee
with 
\be
\xi_{min}~ = ~ x ~{m \over M_A} \quad .
\label{min}
\ee

Then, defining the nucleon structure function $F^N_2(z)$
\be
F^N_2(z) ~ =   -  ~ {1 \over 2} ~ z ~ g_{\mu\nu} ~{1 \over 2} ~\sum_{\sigma} ~ w^{\mu \nu}_{N,\sigma,\sigma}({p},{q})
\label{F2N}
\ee
where
\be
z ~ = ~ {Q^2 \over 2 p \cdot  q}  \quad ,
\label{Bjorken1}
\ee
one obtains

\be
\fl F^A_2(x) =  ~ \sum_N ~ \sumint {d\epsilon }
  \int {d{\bm \kappa}_{\perp} \over (2 \pi)^3} ~\int_{\kappa^+_m}^{\infty} ~ {d\kappa^+ \over 2~ \kappa ^+} ~
    ~ 
 {\cal P}^{N}(\tilde{\bm \kappa},\epsilon) ~ ~ {P^+_A \over  p^+} ~ {Q^2 \over 2 P_A \cdot  q}
~ { 2 p \cdot  q\over  Q^2} ~
 F^N_2(z) ~=
 \nonu
\fl =  ~ \sum_N ~
\sumint {d\epsilon }
  \int {d{\bm \kappa}_{\perp} \over (2 \pi)^3}~ \int_{\kappa^+_m}^{\infty} ~ {d\kappa ^+ ~ \over 2~ \kappa ^+} ~
    ~ 
 {\cal P}^{N}(\tilde{\bm \kappa},\epsilon) ~
 F^N_2(z) ~\quad ,
 \label{F2a}
\ee
since in our reference frame ${\bf P}_{A \perp} = 0$, ${\bf q}_{\perp} = 0$  and, in the Bjorken limit, $q^+$
is vanishing small with respect to $q^-$.

\vspace{1mm} 
Let us stress
that one cannot integrate over {{$\epsilon$}} to obtain the momentum distribution because both $p^+$ and $p^-$
depend on {{$\epsilon$}}.

\section{Preliminary results for the EMC effect in $^3He$}
\label{PR}
We aim to compare the results of our novel approach for the spectral function and  for the structure function $F_2^A(x)$  in the DIS limit with the available experimental data for the EMC effect in $^3He$.
To this end
{{{we use  the  Pisa group $^3He$ wave  function  \cite{Kievsky} corresponding to the AV18 NN interaction \cite{AV18} to 
 obtain the LF spectral function 
 from (\ref{SFex}).} }}
 
 Then we have to evaluate $F_2^A(x)$ from (\ref{F2a}), the ratio
 
  \be
R^A_2(x)={F^A_2(x)\over Z~F^p_2(x)+(A-Z)~F^n_2(x)}
\ee
 for $A$=3 (i.e. for $^3He$) and for $A$=2 (i.e. for the deuteron) and eventually the EMC ratio
  
 \be
r_{EMC}(x)={R^{He}_2(x)\over R^D_2(x)}  \quad .
\ee
\bfi
\vspace{3mm} 
\centerline
{\includegraphics[width=11.50cm,
]{emc_6.eps}}
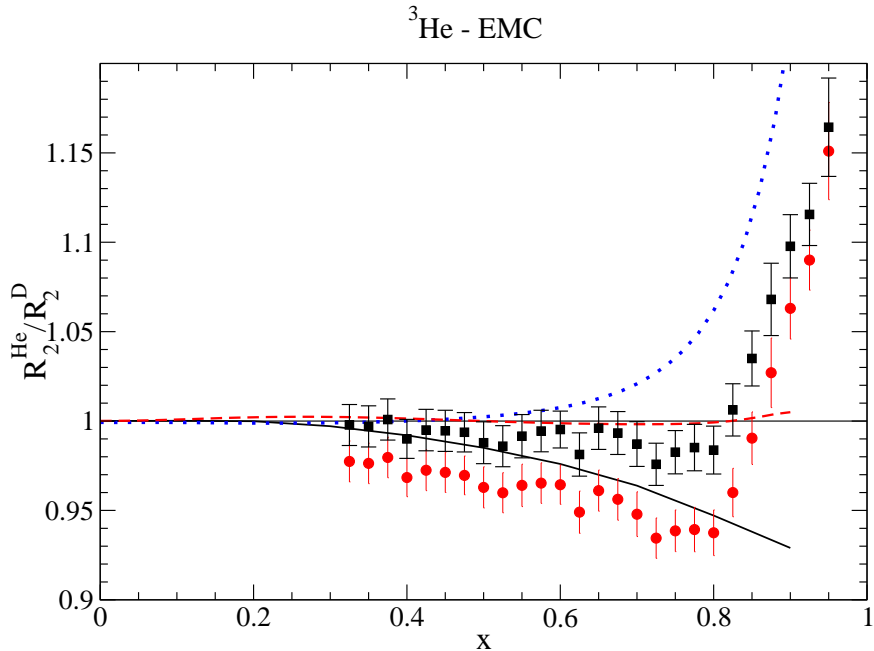
{ \caption{Results for the EMC effect in $^3He$. Solid line: the two-body contribution
obtained with the {{LF Spectral Function}} (see equation (\ref{F2a})), renormalized for a comparison with the experimental data (see text). 
Dotted line : full results obtained  with  the approach of \cite{Sauer} for the spectral function. Dashed line : two-body contribution with the convolution formula  of \cite{Sauer}, properly renormalized.
 All the evaluations are obtained with the AV18 NN interaction \cite{AV18}.
 Red dots : experimental data from \cite{Seely}; black squares reanalysis of the experimental data from \cite{KuPe}.}}
\label{EMC2}
\efi

In Section \ref{SF} it is noticed that for an evaluation of the LF spectral function both the bound and the scattering states in momentum space are needed. Then  the tricky task to calculate  the scattering states  in momentum space has to be solved. As a preliminary calculation, we evaluate in the Bjorken limit only the contribution  from the two-body channel to the EMC effect in $^3He$, i.e. we take from (\ref{F2a}) only  the contribution with the spectator $(A-1)$=2 state in the deuteron state. 
To properly compare with the experimental data the two-body contribution to the EMC ratio, 
this contribution is renormalized as to obtain $r_{EMC}(x=0)=1$. The 
renormalization factor is
1/0.679, 
as expected since  the probability to find a deuteron in $^3He$ is 0.679 by using the AV18 NN interaction.

In Fig. 1, the experimental data of the JLab E03-103 experiment \cite{Seely} (red dots) and the outcomes of  \cite{KuPe} (black squares), where a carefully reanalysis of the same experimental data was carried out, are compared with theoretical calculations. The authors of \cite{KuPe} in their analysis indicate that a normalization factor is required  in the E03-103 data for $^3He$ to be consistent with the $F^n_{2}/F^p_{2}$ ratio from the NMC experiment \cite{NMC}. With this normalization the  E03-103 data for $^3He$ are in good agreement with the HERMES data \cite{HERMES} in the overlap region at 
$x \sim 0.35$.

The solid line is our renormalized two-body contribution to the EMC ratio.
n Fig. 1 the results for $F_2^A(x)$ obtained through a convolution formula  with the approach of \cite{Sauer} for the spectral function are also shown. The dotted line  is the full result, while the dashed line is the contribution from the  two-body channel, properly renormalized. The essential difference of our LF spectral function with respect to the approach of \cite{Sauer} is the use of the intrinsic momentum $\tilde{\bm \kappa}\equiv (\kappa^+,{\bm \kappa}_{\perp})$ of a free nucleon in the reference frame of the  cluster $[1, (A-1)]$, where the system of ~$(A-1)$ nucleons is fully-interacting. 
Noteworthy, the intrinsic momentum $\tilde{\bm \kappa}\equiv (\kappa^+,{\bm \kappa}_{\perp})$  is different from the momentum $\tilde{\bm p} \equiv ( p^+,{\bf p}_{\perp})$ in the laboratory frame, used for the spectral function in \cite{Sauer}, and from the intrinsic momentum in a system of $A$ free nucleons \cite{KP}.

{{It is clear that the LF spectral function pulls down the two-body contribution to the EMC effect  with respect to the approach of \cite{Sauer}, and suggests that a complete calculation including the two- and three-body contributions can be near to the  $^3He$ data,
especially in view of the analysis
 of Kulagin and Petti \cite{KuPe}. Therefore our  next step will be  the full calculation of the EMC effect for $^3He$, including the exact three-body contribution. 
}}
For an approximate evaluation of the three-body contribution see \cite{prel}.

\section{Conclusion and perpectives}
\label{CP}

{{{{A Poincar\'e covariant description of nuclei, based on  the light-front relativistic Hamiltonian
dynamics, has been proposed.}}}}
 {{The Bakamjian-Thomas construction of the Poincar\'e generators 
  allows one to embed the successful phenomenology of interacting nucleons for few-nucleon systems in a
Poincar\'e covariant framework.}}

Our main tool  is the LF spectral function introduced in \cite{noi}, which allows one to satisfy 
the baryon number sum rule and the momentum sum rule at the same time. The definition of LF spectral function is based on the overlaps between the ground state  of the $A$-nucleon system and the  cluster $[1,(A-1)]$, which is the  tensor product of a momentum eigenstate of a free nucleon of momentum {{$\bm \kappa$}} in the intrinsic reference frame of the cluster $[1,(A-1)]$ times the state of  a fully interacting $(A-1)$-nucleon spectator subsystem.
{{The definition of the nucleon momentum  $\bm \kappa$ and the use  
of  the tensor product 
  allows one  to take care of macrocausality and  to introduce 
   {{a novel effect of binding in the spectral function.}}}}
  
The LF spectral function can be obtained from  non-relativistic wave functions  with a realistic nuclear interaction including two- and three-body forces. As shown in Section \ref{FF2}, through the LF spectral function and a current which satisfies Poincar\'e  covariance and current conservation one can write 
an hadronic tensor for inclusive lepton-nucleus scattering to evaluate DIS processes in the impulse approximation. Generalizations are straightforward  : i) to describe exclusive processes \cite{Dupre:2015jha} or semi-inclusive deep inelastic scattering (SIDIS) processes \cite{SIDIS,SIDISfs}; ii) to include the final state interaction between the $(A-1)$-spectator system and the final debris with a Glauber approximation through a distorted LF spectral function as in \cite{SIDISfs};  iii) to finite momentum transfer kinematics  \cite{TBP}.

As a first application of a Poincar\'e covariant description of nuclei, we aim to provide 
a 
calculation of the nuclear part of the EMC effect for  $^3He$ and $^3H$
to unambiguously investigate genuinely QCD-based phenomena. Indeed a Poincar\'e covariant evaluation of this effect  could indicate  the size of the difference with respect to the experimental data to be filled by manifestations of non-nucleonic degrees of freedom or by modifications of nucleon structure in nuclei.

Up to now only  the two-body contribution 
to the LF nucleon spectral function
has been used.
 Encouraging improvements clearly  appear with respect to a  convolution approach with a momentum distribution.
{{{{Our next step will be a full calculation of the three-body contribution.}
} }}

\newpage
\appendix

\section{Orthogonality properties of the momentum eigenstates of the $[1,(A-1)]$ cluster states}
\label{orth}
Let us consider the orthogonality properties of the eigenstates of the $ [1,(A-1)]$ cluster states 
$|{\blf p}\sigma;{\blf P} _{S}; J J_{z} 
\epsilon, \alpha, T_{S} \tau_{S}\rangle_{LF}$ in
the laboratory system
\be
\fl  _{LF}\left\langle  \tau_{S}T_{S} , 
\alpha,\epsilon' J_{z}J;{\blf P}'_S; {\blf p}'\sigma \right|{\blf p}\sigma;{\blf P} _{S}; J J_{z} 
\epsilon, \alpha, T_{S} \tau_{S}\rangle_{LF}
 = \nonu
\fl 2 p^+ (2\pi)^3 \delta^3({\blf p}'-{\blf p})~
 2 P^+_S (2\pi)^3 \delta^3({\blf P}'_S-{\blf P}_S)~ _{LF}\left\langle  \tau_{S}T_{S} , 
\alpha,\epsilon' J_{z}J \right| J J_{z} 
\epsilon, \alpha, T_{S} \tau_{S}\rangle_{LF}
 \label{ortimp1} \ee
 
 Then defining ${\blf P}= {\blf p} + {\blf P}_S$, ${\blf P}'= {\blf p}' + {\blf P}'_S$ and 
 $\xi = p^+/P^+$, $\xi' = p'^+/P'^+$, one has
 \be
\fl  _{LF}\left\langle  \tau_{S}T_{S} , 
\alpha,\epsilon' J_{z}J;{\blf P}'_S; {\blf p}'\sigma \right|{\blf p}\sigma;{\blf P} _{S}; J J_{z} 
\epsilon, \alpha, T_{S} \tau_{S}\rangle_{LF}
= 2 \xi ~ (2\pi)^3~ 
\delta^2({\bf \kappa}'_{\perp}-{\bf \kappa}_{\perp})~
\delta(\xi' - \xi)~
  \nonu
\fl \times  ~ 2(1- \xi) P^+ (2\pi)^3~  \delta^3({\blf P}' - {\blf P})~ _{LF}\left\langle  \tau_{S}T_{S} , 
\alpha,\epsilon' J_{z}J \right| J J_{z} 
\epsilon, \alpha, T_{S} \tau_{S}\rangle_{LF}  \quad ,
 \label{ortimp2} \ee
 where ${\bf \kappa}_{\perp}= {\bf p_{\perp}}-\xi ~{\bf P_\perp}$ and 
${\bf \kappa}'_{\perp}= {\bf p'_{\perp}}-\xi' ~{\bf P'_\perp}$ . 

Introducing 
\be
{\cal{M}}^2_0[1,(A-1)] ~ = ~ {m^2+|{\bm \kappa}_{\perp}|^2 \over \xi} +
{M_S^2+|{\bm \kappa}_{\perp}|^2 \over (1-\xi)} 
\label{M0x}
\ee
and $\kappa^+={\cal{M}}_0[1,(A-1)] ~ \xi $, with the help of (\ref{derparz}) one obtains 
\be
\fl  _{LF}\left\langle  \tau_{S}T_{S} , 
\alpha,\epsilon' J_{z}J;{\blf P}'_S; {\blf p}'\sigma \right|{\blf p}\sigma;{\blf P} _{S}; J J_{z} 
\epsilon, \alpha, T_{S} \tau_{S}\rangle_{LF}
 =
 2 \kappa^+ ~ (2\pi)^3~ 
\delta^2({\blf \kappa}'-{\blf \kappa})~
\nonu \nonu
\fl \times ~ {E_S \over {\cal{M}}_0[1,(A-1)]}~ 2~ P^+ (2\pi)^3~\delta^3({\blf P}' - {\blf P})~ 
  _{LF}\left\langle  \tau_{S}T_{S} , 
\alpha,\epsilon' J_{z}J \right| J J_{z} 
\epsilon, \alpha, T_{S} \tau_{S}\rangle_{LF}   \quad .
\ee
As a consequence one has
\be
|{\blf p}\sigma;{\blf P} _{S}; J J_{z} 
\epsilon, \alpha, T_{S} \tau_{S}\rangle_{LF}
 = 
 \nonu |{\blf P} ={\blf p} + {\blf P} _{S}\rangle_{LF} ~ |{\blf \kappa}\sigma; J J_{z} 
\epsilon, \alpha, T_{S} \tau_{S}\rangle_{LF}~\sqrt{{ E_S \over {\cal{M}}_0[1,(A-1)]}}  \quad .
\label{ortimp5}
 \ee
The factor $\sqrt{{ E_S / {\cal{M}}_0[1,(A-1)]}}$ 
 takes care of the proper normalization
of the momentum eigenstates $|{\blf p} \rangle_{LF}$, $| {\blf P}_{S}\rangle_{LF}$ 
and $|{\blf p} + {\blf P} _{S}\rangle_{LF}$.

\end{document}